\documentclass[11pt]{article}

\usepackage[a4paper,margin=2.5cm]{geometry}
\usepackage[utf8]{inputenc}
\usepackage[T1]{fontenc}
\usepackage{amsmath,amssymb}
\usepackage{graphicx}
\usepackage{float}
\usepackage{cite}
\usepackage{hyperref}

\newcommand{\NLDW}{\mathrm{NLDW}}
\newcommand{\eff}{\mathrm{eff}}

\title{Nonlinear Density Waves and a Galactic-Scale Estimate of the Graviton Mass}
\author{Miroslava Vukcevic\\
Astronomical Observatory Belgrade\\
\texttt{vuk.mira@gmail.com}}
\date{}

\begin{document}
\maketitle

\begin{abstract}
In February 2016 the LIGO and Virgo collaborations reported the discovery of
 gravitational waves in merging black holes, with the conclusion that their observational
 data did not show any violation of general relativity. The joint LIGO and Virgo team
 presented an upper limit on the graviton mass, $m_g < 1.2\times10^{-22}\,\mathrm{eV}$,
 using tests of modified gravitational dynamics. Yukawa-type gravitational potential
 corrections are often used as generic finite-range parametrisation in modified gravity
 contexts, including massive-gravity inspired models. Yukawa corrections are derived by semi-empirical approach, in the framework of $f(R)$ gravity, a type of modified gravity which generalises Einstein’s general relativity. 
Contrary to this approach, we use
 a classical Newtonian gravitational potential corrected by nonlinear effects in order to
 obtain a characteristic wavelength of the nonlinear density wave. This wavelength is then
 phenomenologically identified with an effective graviton Compton wavelength, leading to
 a model-dependent estimate of an effective graviton mass within the galaxy scale.
 The presented approach is independent from other methods published until now.
 \end{abstract}

\section{Introduction}

Since 2016 the LIGO and Virgo collaborations presented the first discovery of gravitational
waves (GW) in merging black holes, together with an upper limit on the graviton mass such as
$m_g < 1.2\times10^{-22}\,\mathrm{eV}$~\cite{Abbott2016a}. Although their observational data
showed no violation of classical general relativity (GR), several perspectives occurred for
cosmology~\cite{Abbott2017}, sources of GW~\cite{Abbott2018}, and testing the standard theory
of gravity in regimes previously out of scope.

Recent simultaneous measurements of arrival times in strongly lensed GW signals and associated
electromagnetic counterparts give an opportunity to estimate GW propagation speed~\cite{Branchesi2016,ColletBacon2017}.
These time delays depend on the matter distribution in the lens galaxy, on the matter distribution
along the line of sight, and on the cosmological parameters~\cite{ColletBacon2017,Fan2017}.

On the other hand, it was recently proposed that dark matter interacting with GW can produce
shear viscosity which can be used as a mechanism to explain the accelerated expansion of the
Universe. However, neither the proposed models nor the standard model of elementary particles
provide a satisfactory explanation for two huge discrepancies between observations and expectations
based on theoretical models. One of many tensions is the inability of the $\Lambda$CDM model to
fully explain dark matter and dark energy phenomena as a complete theory. A number of alternative
models have been proposed in the literature, including those based on modification of the fundamental
laws of gravity: MOND~\cite{Moffat2005,Moffat2020}, nonlocal theories~\cite{Dimitrijevic2020},
or Yukawa-type gravitational-potential corrections~\cite{Will1998,Sanders1986}. The weakness of
many of these theories is that their functional form is often introduced phenomenologically and is
not unique over a wide range of spatial scales.

Other frameworks postulate a nonzero graviton rest mass~\cite{GoldhaberNieto2010,deRham2017},
incorporating this mass into suitable modified theories of gravity. Assigning a finite mass to
the graviton requires a modification of the Newtonian gravitational potential. However, such
modifications must also address major astrophysical evidence commonly associated with dark matter,
including the flat rotation curves of stars in galaxies~\cite{Rubin2000,BertoneHooper2018} and
the stability of galaxy clusters~\cite{Zwicky1933,Smith1936,HagueWilkinson2013}, which suggest
the presence of a large amount of non-baryonic matter.

Without entering into the details of the dark matter problem, we propose a new approach for
estimating an effective graviton mass based on nonlinear galactic dynamics. This model is an
extension of the gravitational potential based on Newtonian gravity, related to the natural
formation of spiral structure due to dispersion balanced by nonlinear effects. The role of marginal
stability is to select the regime where the linear density-wave description becomes insufficient
and where dispersive spreading may be balanced by nonlinearity. This balance is the physical
mechanism behind the mathematical existence of the nonlinear Schr\"odinger equation and the corresponding
soliton-like density wave solution.

\section{Model}

A galaxy is an $N$-body system composed of stars and gas, interacting with each other through a
long-range gravitational force. There are many common characteristics between the physics of
galaxies and plasmas, systems of charged particles interacting through the Lorentz force.
A macroscopic description of these systems can be given by a fluid model with incorporated
Poisson's equation, which includes collective behaviour such as waves or instabilities.

Linear density wave theory by Lin and Shu explains the spiral pattern formed within a disk~\cite{LinShu1964}.
Within the infinitesimally thin disk approximation, a spiral-wave solution has been obtained,
but this theory failed to explain long-lasting structure because differential rotation would wind
up the pattern on a timescale insufficient for observation. There were several attempts to overcome
this problem~\cite{Toomre1969,KormendyNorman1979,Kendall2011,Purcell2011}. Our approach extends
the Lin--Shu model by keeping nonlinear terms in expanded variables. The density wave model is
based on the fluid description of the galactic disk. Namely, there are transport equations for the
mass density $\rho$ and the momentum $\rho v$, together with Poisson's equation that relates the
density to the gravitational potential $\phi$. We keep the same notation as in Ref.~\cite{Vukcevic2014}
for the stellar disk.

The equilibrium state of the system is described as a rotation with angular velocity $\Omega(r)$
about the $z$-axis under the balance of centrifugal, gravitational and pressure forces in a frame
rotating with constant angular velocity $\Omega_0$. Then the equilibrium velocity is
$v_{0\varphi}=(\Omega-\Omega_0)r$, where
\begin{equation}
    \Omega^2 r = -\frac{\partial \phi_0}{\partial r} .
\end{equation}
All quantities with subscript $0$ are reserved for equilibrium functions. Pressure is neglected
for simplicity, since the gas follows almost the same pattern as the stellar component, slightly
inclined to it~\cite{Vukcevic2024}.

The dispersive property originates from the coupled Poisson equation, which is a second-order
elliptic partial differential equation. At the same time, this particular equation defines the
geometry of the system. In the infinitesimally thin disk approximation, Lin and Shu assumed a
delta function for the density in the $z$ direction and approximated Poisson's equation by
\begin{equation}
    \frac{\partial \phi(r,z=0)}{\partial r} = \pm 2\pi iG\sigma ,
    \label{eq:poisson_thin}
\end{equation}
where $\sigma$ is the surface mass density~\cite{LinShu1964}. Then the relation between surface
density and the two-dimensional potential is
\begin{equation}
    \sigma = -\frac{k}{2\pi G}\phi(z=0),
    \label{eq:sigma_phi}
\end{equation}
where the wave number is
\begin{equation}
    k = -\frac{i}{\phi}\frac{\partial \phi}{\partial r} .
\end{equation}

In order to derive a possible nonlinear equation, it is necessary to discuss the parameter regime
defined by the dispersion equation. It determines the unique transformation of coordinates and
expansion of variables within the reductive perturbation method (RPM)~\cite{Taniuti1964}.
Here we do not go into the details of the procedure, since it has already been done, we rather list the
implications of such a calculation.

For a differentially rotating thin stellar disk, the linearised fluid equations together with
Poisson's equation, assuming plane-wave type variation
\begin{equation}
    f=f(r)e^{i(kr+m\varphi-\omega t)},
\end{equation}
lead to the dispersion relation
\begin{equation}
    (\omega-m\Omega)^2 = \kappa^2 - 2\pi G\rho_0 |k|,
    \label{eq:dispersion}
\end{equation}
where $\omega-m\Omega$ is the Doppler-shifted frequency, $m$ is the azimuthal wave number, an
integer representing the number of spiral arms, and $\kappa$ is the epicyclic frequency due to
differential rotation
\begin{equation}
    \kappa^2 = 2\Omega\left(2\Omega+r\frac{d\Omega}{dr}\right).
    \label{eq:kappa}
\end{equation}

\section{Marginal stability and characteristic NLDW length}

Using Eq.~\eqref{eq:dispersion}, the stability parameter is defined by
\begin{equation}
    k_2 = \frac{\kappa^2}{2\pi G\rho_0},
\end{equation}
where $k_2$ is normalised by $\kappa^2/(2\pi G\rho_0)$, so all waves with $k<k_2$ are purely
stable. The initial limitation on the wave number $k>k_1$ regarding vertical stability of the disk,
with
\begin{equation}
    k_1=\max\left\{\frac{1}{r},\frac{\rho_0'(r)}{\rho_0(r)}\right\},
\end{equation}
provides a lower limit of the wavenumber. Observational data suggest that $k_1\approx k_2$ in real
galaxies~\cite{Bertin2000}, defining marginal stability of the galactic disk. In this marginally
stable case, a special transformation of variables has to be introduced, different from the stable
case because the frequency goes to zero and, consequently, the group velocity becomes infinite~\cite{Watanabe1969}.

Using this coordinate transformation together with a perturbation expansion of the dependent
variables, one obtains a single nonlinear evolution equation of the nonlinear Schr\"odinger type.
In the galactic context, a similar derivation was carried out in Refs.~\cite{Vukcevic2014,Vukcevic2024},
and therefore we do not repeat the details here.

For the purpose of estimating an effective graviton mass, the relevant quantity is first the
characteristic length scale of the nonlinear density wave solution. The marginal-stability condition
is expressed in terms of a dimensionless wave number $k\sim 1$. In the present model this condition gives
\begin{equation}
    k=\frac{2\pi G\rho_0}{\kappa^2}\frac{2\pi}{\lambda_{\NLDW}}\approx 1,
    \label{eq:nldw_lambda}
\end{equation}
where $\lambda_{\NLDW}$ denotes the characteristic wavelength of the classical soliton-like nonlinear
density mode. Substituting typical Milky Way parameters,
$\rho_0=4\times10^{-2}\,\mathrm{g/cm^2}$ and
$\kappa=10^{-15}\,\mathrm{s^{-1}}$~\cite{Vukcevic2021},
gives the characteristic length scale
\begin{equation}
    \lambda_{\mathrm{NLDW}}\sim10^{17}\,\mathrm{cm}.
    \label{eq:nldw_length}
\end{equation}
Here, $\lambda_{\mathrm{NLDW}}$ denotes the characteristic scale of the
classical soliton-like density wave mode. In the phenomenological
interpretation adopted here, this scale corresponds to an effective
graviton Compton wavelength,
\begin{equation}
    \lambda_g^{\mathrm{eff}}\equiv\lambda_{\mathrm{NLDW}}.
    \label{eq:lambda_eff}
\end{equation}
This corespondency allows one to associate the NLDW length scale with an
effective graviton mass through the ordinary Compton relation
\begin{equation}
    m_g^{\mathrm{eff}}=\frac{h}{c\lambda_g^{\mathrm{eff}}}.
    \label{eq:mg_eff}
\end{equation}
The resulting mass scale should therefore be understood as a phenomenological,
model-dependent effective quantity. It is motivated by the localised,
soliton-like form of the NLDW correction to the potential and does not imply
a quantization of the galactic density wave itself.
\section{Yukawa potential vs NLDW potential}

There are several modified-gravity models, as alternatives to Newtonian gravity, employed to
overcome fundamental issues related to dark matter when explaining galactic or extragalactic
dynamics. One of them is a Yukawa-type modified theory used to derive gravitational-potential
corrections. Since a Yukawa-type correction is commonly used as an effective parametrisation of
finite-range modifications of gravity, including models associated with a nonzero graviton mass~\cite{Sanders1990},
it provides a useful reference potential for comparison with the NLDW-induced potential correction.
In the present comparison, the Yukawa form is used as a generic finite range reference profile.
Yukawa-like potentials have been obtained in the framework of $f(R)$ gravity as a general feature
of these modified theories~\cite{CardoneCapozziello2011,Iorio2010}. $f(R)$ gravity is a type of
modified gravity that generalises Einstein's general relativity~\cite{Buchdahl1970}. It represents
a family of models, each defined by a different function of the Ricci scalar. In the case of
$f(R)$ gravity, the functional form of $f(R)$ is not uniquely fixed a priori and is often chosen
phenomenologically in order to reproduce observations at different scales.

On the other hand, the NLDW approach does not assume a specific form of the gravitational-potential
correction or of the density perturbation in advance. The correction is obtained by retaining
nonlinear terms in the perturbation expansion of the disk equations. The role of marginal stability
is to identify the regime where the linear density-wave description becomes insufficient and where
dispersive spreading may be balanced by nonlinear effects. This balance is the physical mechanism
behind the emergence of the NLS equation and the corresponding soliton-like density wave solution.

Next, Yukawa theory is not a purely phenomenological term; it can emerge as an effective correction
in the context of extended gravity, but the physical meaning of such corrections needs to be
confirmed at different scales: short distances, the Solar System, spiral galaxies, and galaxy
clusters. However, there is no unique expression for all scales; different functions or parameter
values are needed in order to explain dynamics at different scales. As pointed out by Capozziello
et al.~\cite{Capozziello2014}, galactic dynamics could improve the theory as an intermediate scale.
In that sense, we compare a Yukawa-type potential at the single galaxy scale with the gravitational
potential correction derived by NLDW theory.

Using RPM to estimate nonlinear effects in density-wave theory~\cite{Vukcevic2014}, the gravitational potential
gradient $\partial\phi/\partial r$ is approximated by
\begin{equation}
    \frac{\partial\phi}{\partial r}
    = -r\Omega^2
    +\sum_{n=1}^{\infty}\sum_{m=-\infty}^{\infty}
    2\pi G\epsilon^n \mathrm{Re}\left[\rho^{(n,m)}(\xi,\eta)e^{i(kr-\omega\tau)}\right],
    \label{eq:potential_gradient}
\end{equation}
where the term $r\Omega^2$ comes from the equilibrium property. Here $\Omega$ is angular velocity,
$G$ is the gravitational constant, $\rho$ denotes surface mass density, $\xi$, $\eta$, and $\tau$ are
corresponding stretched coordinates, while $k$ and $\omega$ are wave number and frequency, respectively.

In order to derive the gravitational potential, we use the already derived solution of the nonlinear
Schr\"odinger equation for the surface-density perturbation,
\begin{equation}
    \rho^{(1,1)}(\xi,\eta)
    =\frac{\rho_a e^{i\psi}}
    {\cosh\left(\sqrt{Q/P}\,\rho_a(\xi-P\eta)\right)}.
    \label{eq:rho_sol}
\end{equation}
The wave phase $\psi$ is not relevant since only the real part of Eq.~\eqref{eq:rho_sol} is taken,
and the parameters $P=\kappa/(\pi G\rho_0)=1/V_g$ and $Q=\kappa^3/(\pi G\rho_0)$ are related to
the soliton velocity $V_g$ and width of the soliton; $\rho_a$ is the wave amplitude and $\kappa$ is
the epicyclic frequency.

We substitute the exact solution for the density, Eq.~\eqref{eq:rho_sol}, into the equation for the
potential gradient, Eq.~\eqref{eq:potential_gradient}, and after integration the expression for the
nonlinear gravitational potential reads
\begin{equation}
    \phi(r)=-\frac{\Omega^2 r^2}{2}+\frac{ar}{\cosh b(T-cr)}+C.
    \label{eq:nldw_full}
\end{equation}
All parameters and variables are dimensionless. Returning to original coordinates, the nondimensional
velocity is multiplied by $2\pi G\rho_0/\kappa$, and $T=1=(t+\varphi/\Omega)$ by $\kappa$, where
$\varphi$ is the polar angle. The parameters $a$, $b$, and $c$ are defined by the NLDW scaling,
and for suitable galactic values they can reproduce a flat shape of rotation curves even for radii
of order Mpc~\cite{Vukcevic2021,VukcevicUniverse2022,VukcevicBonJovanovic2025}, without explicit
inclusion of dark matter.

Before proceeding, it is necessary to discuss the integration. The potential is derived up to a
constant. The first term is close to constant because $\Omega$ is a function of radius; according
to Fich et al.~\cite{Fich1989}, the Milky Way approximately satisfies $\Omega\sim1/r$. For other
galaxies, $\Omega(r)$ is derived from the rotation curve, suggesting a similar power-law dependence.
If the angular velocity follows approximately $\Omega(r)\propto1/r$, as suggested by a flat rotation
curve, then the term $-\Omega^2r^2/2$ becomes approximately constant and can be absorbed into the
integration constant. Therefore, the relevant potential correction reads
\begin{equation}
    \phi_{\NLDW}(r)=\frac{ar}{\cosh b(T-cr)}.
    \label{eq:nldw_corr}
\end{equation}

Here we place the NLDW induced correction side by side with a Yukawa-type finite range profile
at the galaxy scale. The purpose of this comparison is limited: it is not used as evidence
that the two potentials are equivalent or that their radial shapes coincide. For the Yukawa profile
we use a standard finite range form, as used by Will~\cite{Will1998} in the context of flat rotation
curves of single spiral galaxies,
\begin{equation}
    \phi_Y(r)=\frac{GM}{r}e^{-r/r_0},
    \label{eq:yukawa_single}
\end{equation}
and compare it qualitatively with the potential correction derived using the NLDW model,
Eq.~\eqref{eq:nldw_corr}. If the full gravitational potential is written with the usual sign
convention, an overall negative sign should be included. In the following figures, only normalized
positive radial profiles are shown.

\begin{figure}[H]
\centering
\includegraphics[height=7cm]{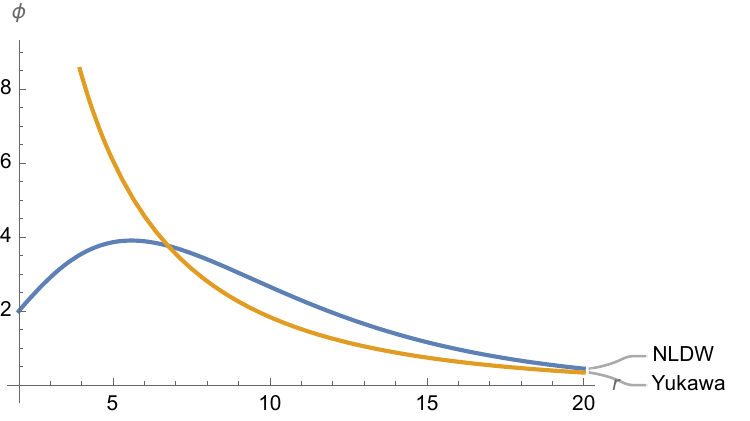}%
\caption{Qualitative comparison of the normalized radial profiles of a Yukawa-type finite-range gravitational correction, Eq.~\eqref{eq:yukawa_single}, and the nonlinear contribution to the gravitational potential, Eq.~\eqref{eq:nldw_corr}, in the range from $1\,\mathrm{kpc}$ to $20\,\mathrm{kpc}$. The NLDW curve corresponds to $a=1$, $b=0.5$, and $c=0.5$, while the Yukawa-type curve corresponds to $GM=50$ and $r_0=10$. The parameters in the two models are chosen independently; therefore, the figure illustrates only a qualitative profile comparison. }
\label{fig:fig1}
\end{figure}

Figure~\ref{fig:fig1} shows the two normalised profiles in the range $(1-20)$ kpc. Since the
parameters in the two models are chosen independently, the curves should not be interpreted as a
fit or as quantitative agreement. The figure is retained only to illustrate that both descriptions
introduce characteristic finite-scale radial contributions in the galactic range. The Yukawa profile
in Eq.~\eqref{eq:yukawa_single} is used here as a generic reference profile, while the
NLDW model is formally applicable from about $1$ kpc upward, where the disk approximation is
appropriate.

\begin{figure}[H]
\centering
\includegraphics[height=7cm]{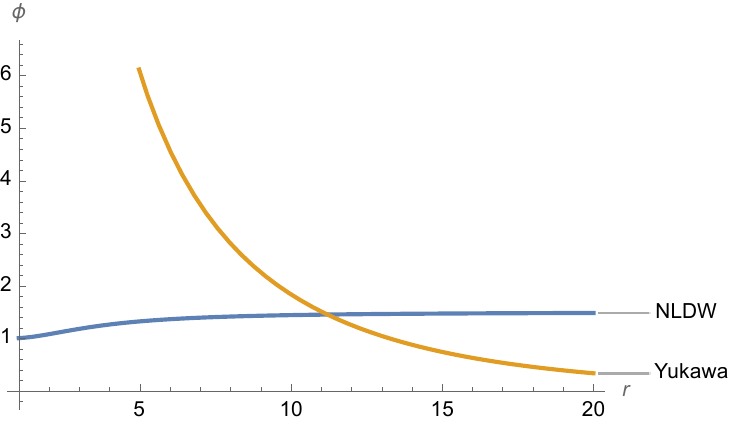}%
\caption{Qualitative comparison of the normalized radial profiles of the Yukawa-type gravitational correction, Eq.~\eqref{eq:yukawa_single}, and the NLDW contribution to the gravitational potential, Eq.~\eqref{eq:nldw_corr}, for a Mestel disk. The radial dependences of $\Omega(r)$, $\rho_0(r)$, and $\kappa(r)$ are taken from Eq.~(7) of Ref.~\cite{VukcevicBonJovanovic2025} and the corresponding NLDW potential.}
\label{fig:fig2}
\end{figure}

Figure~\ref{fig:fig2} repeats the qualitative comparison for an adopted Mestel disk, taking
into account that $\Omega$, $\rho_0$, and $\kappa$ are all $r$ dependent and obey power-law
behaviour~\cite{VukcevicBonJovanovic2025}. This case is closer to the observationally motivated
disk scaling, but it is still a normalised profile comparison. However, NLDW approach provides physical background with no intervention on the mathematical functions used to fit the observational data contrary to  the Yukawa potential.

The NLDW correction and the Yukawa-type correction are derived from different assumptions and, as
the figures indicate, their radial profiles need not be visually similar. The relevant common point
is only that both introduce a characteristic radial scale at galactic distances. The integer $m$
introduced in the model section through the fluid-description ansatz is the classical azimuthal mode
number, counting the number of spiral arms. Thus, the discreteness of the spiral pattern enters
through the integer mode number $m$ in a classical density wave description, not through
quantum-mechanical quantisation of the density wave. This clarification allows the NLDW potential
to be compared phenomenologically to the Yukawa-type one, while the correspondence with a graviton
Compton wavelength is introduced only at the level of an effective length-scale analogy.

Within the NLDW approach, one leaves the assumption of instantaneous two-body interaction in the
strict classical Newtonian sense. Here the relevant dynamics is collective, with most of the mass
placed within the disk. Gravitational interactions have a finite acting time; for example, removing
the black hole from the galactic centre would not result in an instantaneous inertial break-up of
all stars and gas components.

\section{Effective graviton mass}

In quantum-field-theoretic language, the graviton is considered to be the carrier of the
gravitational interaction, a spin-2 tensor boson, electrically uncharged. In general relativity it
is massless and propagates at the speed of light. According to some alternative theories, gravity
may be propagated by a massive field, i.e. by a graviton with a small nonzero mass $m_g$, first
introduced by Fierz and Pauli~\cite{FierzPauli1939}. In the present work, however, the phrase
``effective graviton mass'' is used in a phenomenological sense. It does not denote quantisation
of the galactic density wave itself, but a mass scale obtained by identifying the characteristic
length of the classical soliton-like density mode with an effective Compton wavelength.

Several current theoretical models of massive gravity have established values or bounds, for example:
\begin{itemize}
\item Goldhaber and Nieto found $m_g<2\times10^{-62}\,\mathrm{g}=1.1\times10^{-29}\,\mathrm{eV}$ based on the assumption of a Compton wavelength $\lambda_g=580\,\mathrm{kpc}$, a typical distance between galaxies~\cite{GoldhaberNieto2010};
\item Gershtein et al. established the graviton mass to be $m_g<(1.3-3.2)\times10^{-66}\,\mathrm{g}$ based on a $\Lambda$CDM cosmological model~\cite{Gershtein2003};
\item Will obtained $m_g<7.2\times10^{-23}\,\mathrm{eV}$ at the $2\sigma$ level using a Yukawa potential in the vicinity of the Solar System~\cite{Will1998};
\item Choudhury et al. derived a Compton wavelength $\lambda_g>100\,\mathrm{Mpc}=3\times10^{21}\,\mathrm{km}$ by analysing weak gravitational lensing data, and consequently $m_g<6\times10^{-32}\,\mathrm{eV}$~\cite{Choudhury2004};
\item Finn and Sutton evaluated $m_g<7.6\times10^{-20}\,\mathrm{eV}$ based on binary-pulsar data~\cite{FinnSutton2002}.
\end{itemize}

Some massive gravity theories have been discussed as possible alternatives to the dark energy
explanation of the accelerated expansion of the Universe. In such theories, one important prediction
is that gravitational waves may exhibit frequency-dependent propagation speed. Thus, the effective
gravitational potential may include a correction depending on the Compton wavelength of the graviton,
$\lambda_g=h/(m_gc)$. In other words, if gravitation is propagated by a massive field, then the
effective Newtonian potential may have, for example, a Yukawa form proportional to
$r^{-1}e^{-r/\lambda_g}$. Then the massive graviton propagates at an energy-dependent speed,
\begin{equation}
    \frac{v_g^2}{c^2}=1-\frac{m_g^2c^4}{E^2}
    =1-\frac{h^2c^2}{\lambda_g^2E^2},
\end{equation}
or, equivalently, for frequency $f$,
\begin{equation}
    \frac{v_g^2}{c^2}=1-\frac{c^2}{f^2\lambda_g^2}.
\end{equation}

The above modified dispersion relation is obtained from the modified special relativistic relation
between energy $E$ and momentum $p$ of the graviton,
\begin{equation}
    E^2=p^2c^2+m_g^2c^4,
\end{equation}
with its velocity satisfying $v_g/c=cp/E$. Using the time difference between GW and electromagnetic
waves emitted from the same object, $\Delta t=\Delta t_a-(1+z)\Delta t_e$, where $\Delta t_a$ and
$\Delta t_e$ are differences in arrival and emission times of the two signals, and $z$ is the redshift
of the object, one can obtain an observational constraint on the $v_g/c$ ratio. Using $\Delta t$, the
speed difference reads
\begin{equation}
    1-\frac{v_g}{c}=5\times10^{-17}\left(\frac{200\,\mathrm{Mpc}}{D}\right)
    \left(\frac{\Delta t}{1\,\mathrm{s}}\right),
\end{equation}
where $D$ is the distance of the object~\cite{Will1998,Will2014}.

Here we present an example of Yukawa gravity theory and its application to constraints on the
Compton wavelength and graviton mass; it is used only as a comparison to the NLDW model. In the
general Yukawa-like potential of the form
\begin{equation}
    \phi(r)=\frac{G_{\infty}M}{r}\left(1+\alpha e^{-r/r_0}\right),
\end{equation}
the interaction strength $\alpha$ is not unique but rather defines the scale of interaction, so the
gravitational constant measured at infinity $G_{\infty}$ and locally $G_0$ are related by
$G_0=G_{\infty}(1+\alpha)$, while $r_0$ is the characteristic length scale. If the length scale
$r_0$ corresponds to a graviton mass $m_0$ as $r_0=h/(m_0c)$, then flat rotation curves of spiral
galaxies could be accounted for with $\alpha\sim-1$ without introducing a dark matter hypothesis~\cite{Sanders1986}.
The negative sign of the Yukawa strength indicates an additional repulsive contribution, which
could mimic dark-energy effects on larger scales. However, the weakness of this theory is the fact
that the correction $\alpha$ has to change its value for short scales, such as Solar-System or close
binary-system scales, and long scales, such as galaxies or galaxy clusters.

Using the analogy with the Yukawa potential, we associate the characteristic length scale obtained
from the NLDW theory with an effective graviton Compton wavelength. Since the NLDW model gives
$\lambda_g^{\eff}\equiv\lambda_{\NLDW}\sim10^{17}\,\mathrm{cm}$, the ordinary Compton relation
$\lambda_g^{\eff}=h/(m_g^{\eff}c)$ gives
\begin{equation}
    m_g^{\eff}\sim1.2\times10^{-21}\,\mathrm{eV}/c^2 .
\end{equation}
This value should be understood as a model dependent estimate obtained from galactic scale nonlinear
density wave dynamics. It is not intended to represent a competitive upper bound in the same sense
as constraints derived from galaxy cluster or Solar System tests.

\section{Discussion}

The combination of gravitational wave observations and dynamical tests provides important constraints
on possible deviations from general relativity, including bounds on a nonzero graviton mass. In
gravitational wave analyses, the graviton mass is usually constrained through the modified dispersion
relation for gravitational waves, while Yukawa-type corrections are more naturally connected with
static or weak-field limits of massive gravity inspired models.

LIGO and Virgo collaborations used modified propagation and finite-range considerations to compare
static and dynamical bounds on the graviton mass~\cite{LIGOVirgo2016a,LIGOVirgo2016b}. Static bounds,
such as those from Solar-System observations, do not directly probe the propagation of gravitational
interaction, while dynamical bounds follow very massive and compact objects, such as binary pulsars
and supermassive Kerr black holes. A dynamical bound from GW150914 corresponds to a graviton mass
of order $m_g<1.2\times10^{-22}\,\mathrm{eV}$ at $90\%$ confidence~\cite{LIGOVirgo2016a}. A bound of
$m_g<2.9\times10^{-21}\,\mathrm{eV}$ has been derived using simulations of the S2-star orbit at the
Galactic Center~\cite{Zakharov2016}. Galaxy clusters give much stronger bounds, $m_g<10^{-29}\,\mathrm{eV}$,
with corresponding $\lambda_g>10^{19}\,\mathrm{km}$~\cite{GuptaDesai2018}. Estimates based on observations
of dynamical properties of astrophysical objects at large scales provide some of the strongest bounds
on the graviton mass compared with measurements of the GW speed~\cite{Abbott2016a,Branchesi2016,Will1998}.

The approach considered in this work is different from a purely phenomenological Yukawa parametrisation.
The characteristic scale is not introduced as a fitting parameter, but it is rather related to the nonlinear
dynamics of spiral density waves and to the marginal stability condition of the galactic disk. Therefore,
the obtained value of the effective graviton mass should be regarded as a model-dependent estimate with
a specific dynamical origin within the NLDW framework.

This estimate does not replace existing bounds obtained from gravitational-wave observations,
Solar System dynamics, binary pulsars, or galaxy cluster studies. Rather, it provides an additional
galactic scale perspective suggesting that nonlinear density wave dynamics may be relevant when
discussing effective length scales associated with modified gravitational potentials. Although the
approach is model dependent, it differs from a purely fitting-based Yukawa parametrisation because the
characteristic length scale is connected to the marginal stability condition of the galactic disk.

We propose an estimate of the effective graviton mass obtained by identifying the NLDW characteristic
length scale with an effective Compton wavelength. This should be interpreted as a phenomenological
length-scale analogy, not as a direct quantisation of the density wave and not as evidence of quantitative
agreement with gravitational wave bounds. The NLDW correction and the Yukawa-type correction used for comparison, both introduce characteristic galactic length scales, but their radial profiles are generally different.
Thus, the comparison supports only a phenomenological analogy, not an equivalence between the two
potentials.

The result based on the nonlinear approach needs to be improved; in particular, it may be complemented
by nonlinear effects incorporated into larger-scale dynamics. This will be the subject of further research.

\section*{Acknowledgements}
Part of this research is supported by the Ministry of Education and Science of the Republic of Serbia
(contract 451-03-66/2024-03/200002).

\end{document}